\documentclass[amssymb,floatfix,pre,aps,notitlepage,twocolumn,showkeys,showpacs]{revtex4-1}

\usepackage{lipsum}
\usepackage{graphicx}%  needed for figures
\usepackage{amsmath}% needed for \tfrac, \bmatrix, etc.
\usepackage{amsfonts}% needed for bold Greek, Fraktur, and blackboard bold need
\usepackage{bm}%
\usepackage{epsfig}%
\usepackage{multirow}%
\usepackage{tabularx}
\usepackage{dsfont}
\usepackage[utf8x]{inputenc}                                                    
\usepackage[english]{babel} 
\usepackage{xcolor}
\usepackage{bbm}
\usepackage{boldline}

\AtBeginDocument{

}
\newcolumntype{Y}{>{\centering\arraybackslash}X}

\begin{document}
\title{Work done on a single-particle gas during an adiabatic compression/expansion process}
\author{Carlos E. Álvarez}
\email{E-mail: carlosedu.alvarez@urosario.edu.co}
\author{Nicolás Afanador}
\affiliation{Departamento de Matemáticas Aplicadas y Ciencias de la Computación, Universidad del Rosario, Bogotá, Colombia}
\author{Gabriel Téllez}
\affiliation{Departamento de Física, Universidad de los Andes, Bogotá, Colombia}

\date{\today}

\begin{abstract}
  We compute the average work done by an external agent, driving a piston at constant speed, over a single particle gas going through an adiabatic compression and expansion process. To do so, we get the analytical expression relating the number of collisions between the piston and the particle with the position of the piston during the process. The ergodicity breaking of the system during the process is identified as the source of its irreversibility. In addition, we observe that by using particular initial distributions for the state of the particle, it is possible to preclude the possibility of a net energy transfer from the agent to the particle during the process.
\end{abstract}

\maketitle

%\linenumbers

\section{Introduction}
\label{sec:intro}

  The origins of the irreversibility of macroscopic systems, whose components follow the reversible laws of mechanics, are an open topic of discussion \cite{Coveney1988,Lieb1999,Jarzynski2011,Uffink2015,Lucia2016,Castelvecchi2017}. Undoubtedly, one of the factors that make the topic so complex is the description of the degrees of freedom of the large amount of particles that compose a macroscopic system. Our aim here is to look for the sources of irreversibility in a simple system, where the microscopic state is easily tractable.
%\refB{
  The idea of using the advantage of tractability of single particle gases follows from the original study of a single particle engine by Szilard, which showed that there is an important relation between information and entropy \cite{Parrondo2001,Leff2003}.\\
%}
  
However, before addressing this question it is important to clarify what is meant by {\it reversibility}. We follow Jaynes \cite{Jaynes1984}, who identifies three different meanings of the term, as used in the literature, for a process that takes the system from state $A$ to state $B$.\\
%These are: {\it (1) Mechanical Reversibility: Reversing all molecular velocities in $B$, the equations of motion carry the system back along exactly its previous path to $A$. (2) Carnot Reversibility: The macroscopic physical process can be made to proceed in the opposite direction $B\rightarrow A$, restoring the original macrostate.} And finally {\it (3) Thermodynamic Reversibility: Even if the backward process $B\rightarrow A$ cannot be made to take place reversibly [...], if by any means such as $B\rightarrow C\rightarrow D\rightarrow A$ the original macrostate can be recovered without external change, then all entropies are unchanged and the process $A\rightarrow B$ is thermodynamically reversible.} This last definition has also been called {\it recoverability} \cite{Uffink2001}.\\

%\refB{
  The three meanings identified by Jaynes refer to different quantities:
%}
Mechanical (or microscopic) reversibility is concerned with the exact microstate of the particles in the system (positions and momenta), and it is satisfied by any Hamiltonian system where the reversal of all the momenta leads to a trajectory which is the exact opposite of the original one. Carnot reversibility has to do with quasistatic processes where the system is ergodic, driven by a set of external parameters that change infinitely slowly with time, and the microscopic trajectories of individual particles are not important as long as the trajectory of the macroscopic variables is reversed. Finally, thermodynamic reversibility only cares for the initial and final macroscopic states of the system and the environment.\\ 

%\refB{
There are previous studies of adiabatic processes on many-paricle gasses enclosed in a container with a piston \cite{Gruber2003,Cencini2007,Gislason2010}.
%}
In taking as a system a single particle and a piston, we pretend to avoid the difficulty (or impossibility) of obtaining a precise description of the microscopic state of the system \cite{Norton2017}, allowing us to track its evolution and identify the sources of irreversibility in such a system, when it goes through a compression/expansion process.\\  

This paper is organized as follows: In section \ref{sec:model} we present the model of the system and describe the processes to be carried out on it. From the definition of the process an analytical solution is found for the position at which the $n$th collision between the piston and the particle takes place, as a function of the number of collisions ($n$) and the parameters of the model. Section \ref{sec:phasespace} presents the analysis of the phase space of microscopic states of the system in terms of the transitions between the macroscopic states. The relation between the location of the $n$th collision and $n$ found in section \ref{sec:model} is used to find the boundaries between regions that, after a compression or expansion process, lead to different final macroscopic states given by the volume of the system and the kinetic energy of the particle. In section \ref{sec:work} the average work done on the system during a compression/expansion process is computed by assuming a particular initial distribution of microstates, and using the phase space diagrams obtained in section \ref{sec:phasespace} to compute the probability of doing a certain amount of work during the process. In section \ref{sec:reversibility} we define a measure of the irreversibility of a path in phase space for the compression/expansion process.
Finally, section \ref{sec:conclusions} presents a summary and discussion of the results.

\section{Model}\label{Model}
\label{sec:model}

The level of coarse graining in the description of a system is important as it represents the state of our knowledge about it\cite{Lindgren2015,Perez2016}. In this study the state of the system will be seen from two coarse graining perspectives: The {\it microscopic} state, which is given by the position and velocity of the particle $(z,v)$, and the {\it macroscopic} state, given by the kinetic energy of the particle and the volume within which it can be found at a given time $(K,L)$.\\

The system consists of a cylinder of length $L_r$ with adiabatic walls and a piston at one of its ends, containing a single particle of mass $\mu$. The coordinate axis $\bm{\hat{z}}$ is oriented along the symmetry axis of the cylinder with the origin ($z=0$) at the left wall. The piston is located initially at $z=L_r$ and can move from $L_r$ to $L_l$ ($0<L_l<L_r$) and back.\\

\begin{figure}[h]
  \centering
  \includegraphics[scale=0.35]{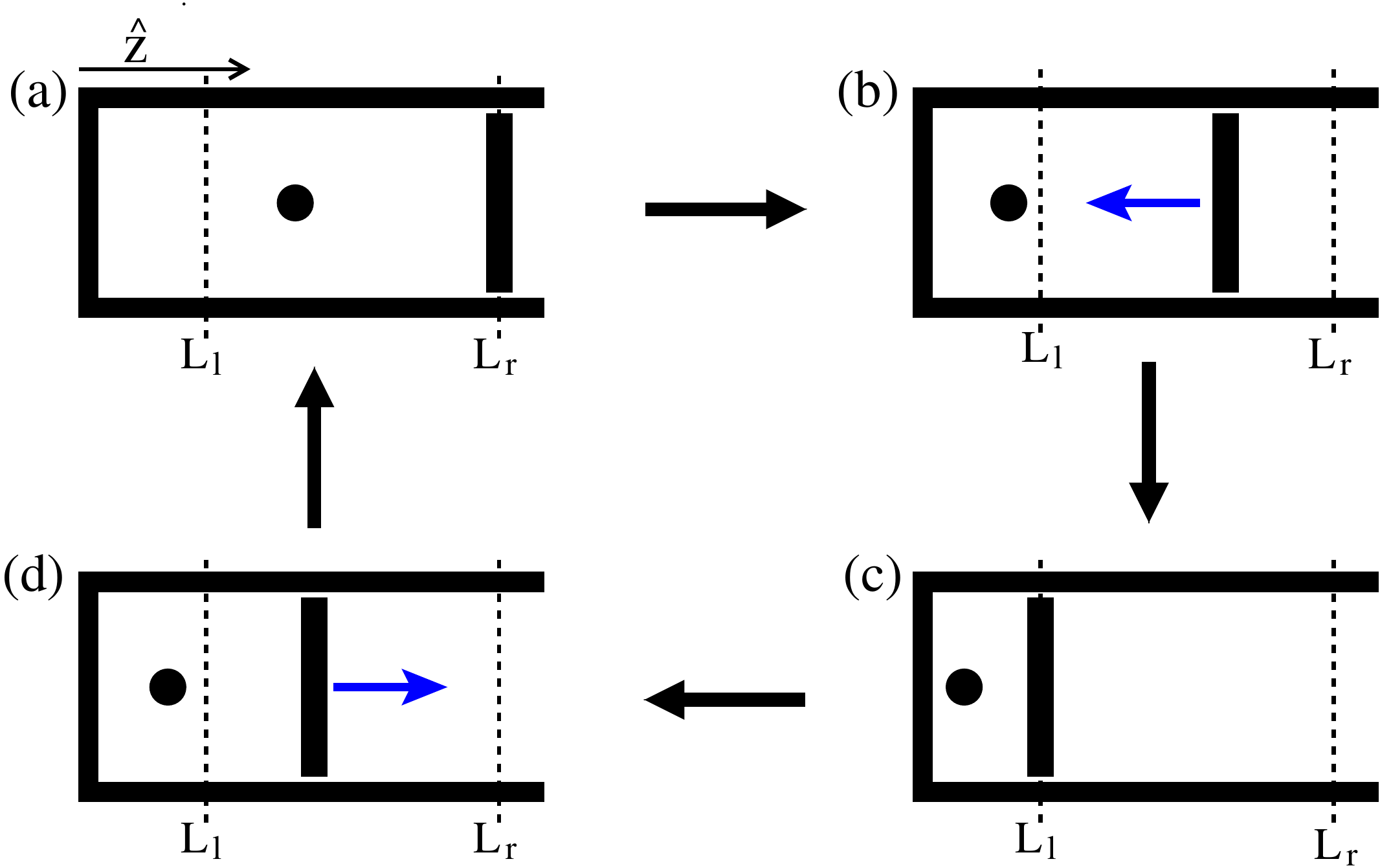}
  \caption{Adiabatic cycle. From an initial position $L_r$ (a) the piston starts compressing the system at constant speed $|u|$ (b), until it stops at $L_l$ (c). After an certain time, the piston starts expanding the system at the same constant speed (d) until it stops back at $L_r$.}
  \label{syst}
\end{figure}

An adiabatic process, in the sense that there is no heat transfer through the system's boundaries,
is carried out (Fig. \ref{syst}) starting from a configuration in which the particle is located at $z$, moving with velocity $\bm{v}$, while the piston, located at $L_r$, moves with a constant velocity $u\bm{\hat{z}}$. As there is no coupling between the movement of the particle in the $\bm{\hat{z}}$ direction and the direction perpendicular to this axis we will treat the system as one-dimensional with the particle moving with velocity $v\bm{\hat{z}}$. At $t=0$ the piston moves to the left and compresses the system adiabatically until it stops at $L_l$ at time $t=(L_r-L_l)/|u|$. At time $t+\Delta t>t$ the piston starts moving again, this time to the right, expanding the system until it gets back to $L_r$ at time $t+\Delta t+(L_r-L_l)/|u|$. Here, $\Delta t$ represents a lapse long enough for the position of the particle to relax to the uniform distribution.\\

We will proceed to analyze the motion of a point particle along
a compression-expansion process as it collides with the piston and the
walls of the container. As we want to look for the possibility of
extracting work from the particle through the adiabatic cycle, we will restrict the analysis to the cases where the initial speed of the particle is greater than that of the piston ($|v|>|u|$), as otherwise it is only possible either to do no net work on the particle or to increase the particle's kinetic energy.\\

The particle and the piston interact through a conservative, short range, repulsive  pair potential $U$. The average force exerted by the piston on the particle during the lapse $\Delta \tau$ in which they interact is
%\begin{linenomath}
\begin{align}
  F=\frac{1}{\Delta \tau}\int_0^{\Delta \tau}\left(-\frac{dU}{dz}\right)dt=\mu\frac{\Delta v}{\Delta \tau},
\end{align}
%\end{linenomath}
where the conservation of momentum, taking into account that the speed of the piston is constant, gives us the change in the velocity of the particle from $v$ to $v'$
%\begin{linenomath}
\begin{align}
  \Delta v=v'-v \label{momCons}.
\end{align}
%\end{linenomath}

When the particle with velocity $v$ collides with the piston, the piston does an amount $W$ of work on the particle.
In order to maintain the velocity of the piston constant, an external agent applies an average force $F$ on the piston each time a collision occurs,
thus doing in turn an amount of work $W$ on the piston. Then, from the conservation of energy we have
%\begin{linenomath}
\begin{align}
  \frac{1}{2}\mu(v'^2-v^2)&=W\nonumber\\
  &=F\Delta z\nonumber\\
  &=\mu\ u\Delta v,\label{enCons}
\end{align}
%\end{linenomath}
where $\Delta z$ is the displacement of the piston during its interaction with both the particle and the external agent, and $u=\Delta z/\Delta\tau$. From equations (\ref{momCons}) and (\ref{enCons}) we obtain the change in the velocity of the particle
%\begin{linenomath}
\begin{align}
  \Delta v=2(u-v),
\end{align}
%\end{linenomath}
and the work done in changing the particle's velocity \cite{Hoppenau2013}
%\begin{linenomath}
\begin{align}
  W=2\mu\ u(u-v).
  \label{work1}
\end{align}
%\end{linenomath}
Additionally, at every elastic collision with the left wall the particle will invert its velocity. From equation (\ref{work1}) we see that if $W>0$, energy is being injected to the system, whereas $W<0$ means that energy is being extracted from the system.\\

As the velocity of the particle will change in a discrete manner after each collision, during the following derivation we will label the velocity as $v_i$, where $i$ indicates the number of hits that have occurred
between the particle and the piston
since the beginning of the process,
and $z_0$ will indicate the position of the particle at the start of the process.
In computing the location $L_1$ at which the first collision between the particle and the piston occurs during the compression ($u=-|u|$, with the piston starting at $L_r$), we have to take into account if the particle starts moving to the left or to the right. In the first case ($v_0>0$) the collision occurs at
%\begin{linenomath}
\begin{align}
  L_1&=\frac{L_r|v_0|+z_0|u|}{|v_0|+|u|},\label{lcomp1}
\end{align}
%\end{linenomath}
provided that it happens before the piston gets to $L_l$, that is, if
%\begin{linenomath}
\begin{align}
  |v_0|>|u|\frac{L_l-z_0}{L_r-L_l}.
\end{align}
%\end{linenomath}
In the second case ($v_0<0$) the particle bounces back at the wall at $z=0$ and then collides with the piston at
%\begin{linenomath}
\begin{align}
  L_1&=\frac{L_r|v_0|-z_0|u|}{|v_0|+|u|},\label{lcomp2}
\end{align}
%\end{linenomath}
if
%\begin{linenomath}
\begin{align}
  |v_0|>|u|\frac{L_l+z_0}{L_r-L_l}.
\end{align}
%\end{linenomath}

In the case of the expansion ($u=|u|$, with the piston starting at $L_l$), if the particle starts moving to the right, the first collision happens at
%\begin{linenomath}
\begin{align}
  L_1&=\frac{L_l|v_0|-z_0|u|}{|v_0|-|u|},\label{lexp1}
\end{align}
%\end{linenomath}
if
%\begin{linenomath}
\begin{align}
  |v_0|>|u|\frac{L_r-z_0}{L_r-L_l}.
\end{align}
%\end{linenomath}
On the other hand, if it starts moving to the left
%\begin{linenomath}
\begin{align}
  L_1&=\frac{L_l|v_0|+z_0|u|}{|v_0|-|u|},\label{lexp2}
\end{align}
%\end{linenomath}
provided that
%\begin{linenomath}
\begin{align}
  |v_0|>|u|\frac{L_r+z_0}{L_r-L_l}.
\end{align}
%\end{linenomath}

For subsequent hits $i$ after the first one ($i>1$) we might consider the particle as starting in a state immediately after the collision, located at the position in which the collision happened ($L_i$).
The velocity of the particle after collision $i+1$ ($v_{i+1}$) in terms of $v_i$ is
%\begin{linenomath}
\begin{align}
  v_{i+1}=|v_i|+2(u-|v_i|)=2u-|v_i|,
  \label{velchange}
\end{align}
%\end{linenomath}
where the $i$ indicates that $i-1$ collisions have occurred before the piston gets to $L_i$.
Notice that during the expansion, whenever $v_{i+1}>0$ the particle will continue to follow the piston without further collisions until after the expansion ends. Otherwise it will move away from the piston and bounce against the wall. The time it takes the particle to go from $L_i$ to the wall at $z=0$ and back to the next collision location $L_{i+1}$, at any step $i>0$ is
%\begin{linenomath}
\begin{align}
  \Delta t_i=\frac{L_i+L_{i+1}}{|v_i|}=\frac{2L_i+u\Delta t_i}{|v_i|},
\end{align}
%\end{linenomath}
%\refA{
which along with $L_{i+1}=L_i+u\Delta t_i$, gives
%}
%\begin{linenomath}
\begin{align}
  L_{i+1}=L_i\frac{|v_i|+u}{|v_i|-u}.
  \label{Lrecurs}
\end{align}
%\end{linenomath}
Supposing now that the current process (compression or expansion) has not ended and $u$ has not changed its sign, equation (\ref{Lrecurs}) implies
%\begin{linenomath}
\begin{align}
  L_{n}=L_1\prod_{i=1}^{n-1}\frac{|v_{i}|+u}{|v_{i}|-u},\ \ i>1.
  \label{len}
\end{align}
%\end{linenomath}
Applying equation (\ref{velchange}) recursively for $|v_{0}|\geq 2iu$ we get
%\begin{linenomath}
\begin{align}
  |v_{i}|=|v_0|-2iu,
  \label{spabs}
\end{align}
%\end{linenomath}
which means that the speed of the particle increases/decreases by $2|u|$ at every collision during the compression/expansion process.\\

Using equation (\ref{spabs}) in (\ref{len}) we obtain
%\begin{linenomath}
\begin{align}
  L_{n}&=L_1\prod_{i=1}^{n-1}\frac{|v_0|-(2i-1)u}{|v_0|-(2i+1)u}\nonumber\\
  &=L_1\frac{|v_0|-u}{|v_0|-(2n-1)u}.
  \label{Lk}
\end{align}
%\end{linenomath}
which gives the position of the piston just as the $n$th  collision occurs during either a compression or expansion process. The value of $L_1$ is obtained from one of the equations (\ref{lcomp1}), (\ref{lcomp2}), (\ref{lexp1}) or (\ref{lexp2}), depending on the initial state of the particle.\\

\section{Phase space}
\label{sec:phasespace}

\begin{figure}[h]
  \centering
  \includegraphics[scale=0.23]{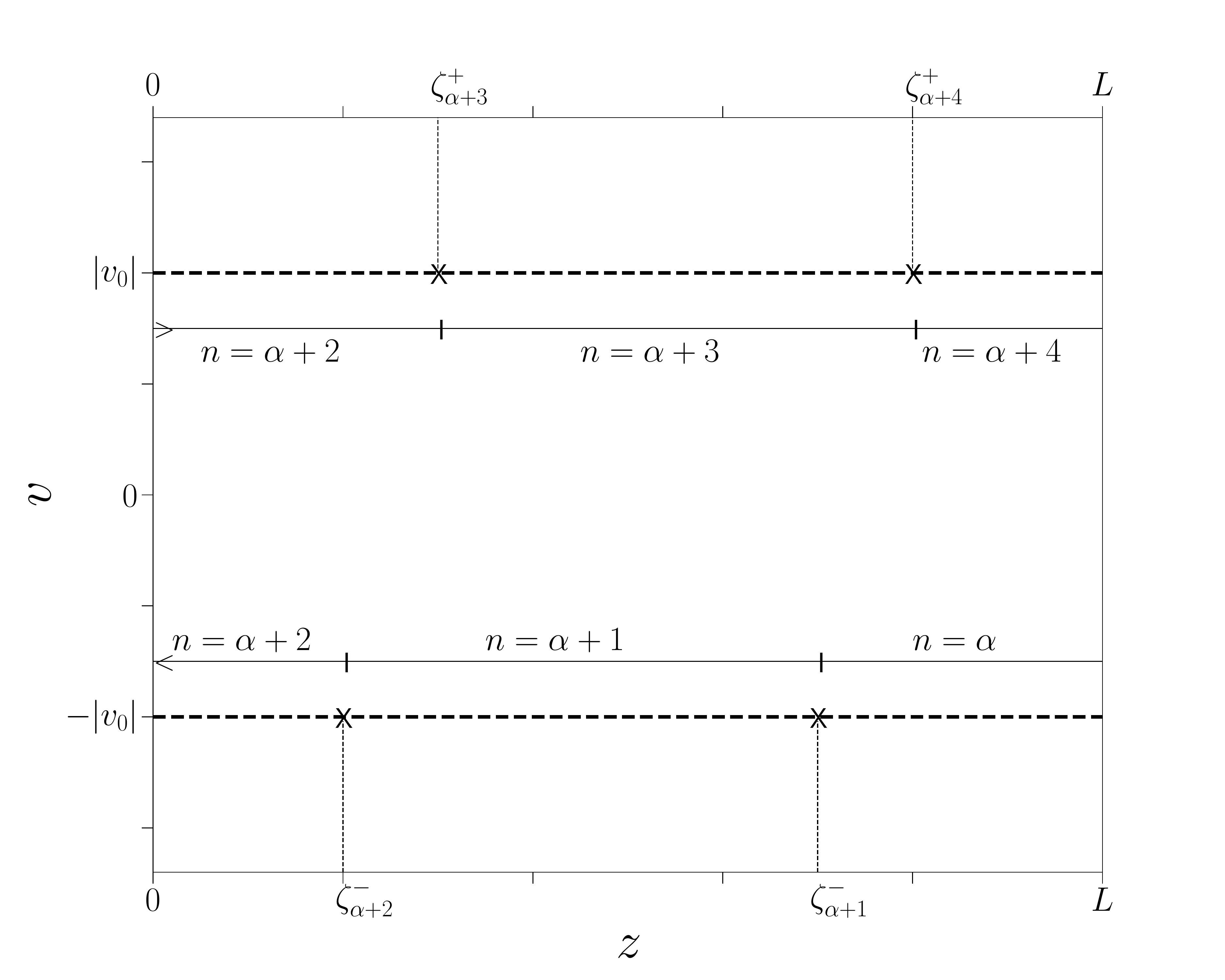}
  \caption{Schematic example of a $v$ vs. $z$
    phase diagram. The region in phase space that can be occupied by the particle at the start of the process ({\bf -\hspace{0.05cm}-\hspace{0.05cm}-}) is divided into regions, each one leading to a different number $n$ of hits occurring (and thus the work done) by the end of the process. The points $\zeta_{n}$ mark boundaries between a region in which $n-1$ hits occur, and another one in which $n$ hits occur. The $+$ or $-$ superscript indicates weather the point is located on the positive or negative velocity region, respectively. $\alpha$ represents the minimum value of $n$ and depends on the parameters of the system.}
  \label{phdtempl}
\end{figure}

Equation (\ref{Lk}) tells us, for a given set of the parameters of the process ($u$, $L_r$ and $L_l$), the location of the $n$th collision as a function of the initial state of the particle and the number $n$ of collisions. As $n$ is an integer, the phase space of the initial state of the particle can be divided into a number $M_c+1$ of regions, each one containing a set of initial
microscopic states $(z,v)$ for which the number of collisions (and
thus the work performed on the system) during an entire compression or
expansion process is equal. To find the limits of such regions and
construct a diagram of the phase space we need to find the points
$(z=\zeta_n,v=v_n)$ at which the number of hits changes from $n-1$ to
$n$ while increasing(decreasing) $z$ when $v$ is positive(negative). Figure \ref{phdtempl} presents a schematic example of the type of phase diagrams we will present further ahead.\\  

To find such points in the case of the compression ($u=-|u|$) we realize that the final hit of the compression process, which occurs at $L_{n}$ after $n$ hits, satisfies
%\begin{linenomath}
\begin{align}
  L_{n}\geq L_l > L_{n+1},
  \label{comp_ineq}
\end{align}
%\end{linenomath}
with $n=\alpha,\cdots,\alpha+M_c$, where $\alpha$ is the minimum
  number of collisions that can happen during the compression. This, along with equation (\ref{Lk}), gives
%\begin{linenomath}
\begin{align}
    f(v_0)<n\leq f(v_0)+1,
  \label{n_ineq}
\end{align}
%\end{linenomath}
where
%\begin{linenomath}
\begin{align}
  f(v_0)=\frac{1}{2}\left(\frac{L_1}{L_l}-1\right)\frac{|v_0|+|u|}{|u|}.
\end{align}
%\end{linenomath}

In case the particle starts moving to the left ($v_0=-|v_0|$), $L_1$
is given by equation (\ref{lcomp2}). At the moment of the last
collision before the compression ends $n$ hits would have occurred,
and
%\begin{linenomath}
\begin{align}
  \zeta_n^{-}=(L_r-L_l)\frac{|v_0|}{|u|}-L_l(2n-1),
  \label{zminus}
\end{align}
%\end{linenomath}
where $\zeta_n^{-}$ denotes a transition point in the region of the phase space in which $v_0=-|v_0|$. In a similar way, if $v_0=|v_0|$ then $L_1$ is given by (\ref{lcomp1}) and a transition point in the positive region is given by
%\begin{linenomath}
\begin{align}
  \zeta_n^{+}=-(L_r-L_l)\frac{|v_0|}{|u|}+L_l(2n-1).
  \label{zminus}
\end{align}
%\end{linenomath}

In order to find all the values for $\zeta_n^{-}$ and $\zeta_n^{+}$ given $L_r$, $L_l$, $|v_0|$ and $|u|$, we start by finding the minimum number of collisions that can happen during the process by placing the particle at $L_r=\zeta_{\alpha}^{-}$, so that
%\begin{linenomath}
\begin{align}
  \alpha=\mbox{int}\left[\frac{1}{2}\left(\frac{L_r}{L_l}-1\right)\left(\frac{|v_0|}{|u|}-1\right)\right],
\end{align}
%\end{linenomath}
where int$[]$ means the integer part. From there it suffices to increase the value to $n=\alpha+1$, $n=\alpha+2$, up to some $n=\alpha+l$ such that the value obtained for $\zeta_{\alpha+l+1}^{-}$ is less than zero, which is out of the container and therefore no longer a valid value for $n$. Then, the values for $\zeta_n^{+}$ are computed starting from $n=\alpha+l+1$, while $0\leq\zeta_n^{+}\leq L_r$.\\

In the case of the expansion ($u=|u|$), given that the compression started at a region where $n$ hits occur (that we will simply call {\it region $n$}), the particle starts at some state $(z_0,w_{0,n})$, where $0<z_0<L_l$ and, by equation (\ref{spabs})
%\begin{linenomath}
\begin{align}
  |w_{0,n}|=|v_0|+2n|u|.
  \label{viexp}
\end{align}
%\end{linenomath}
The final hit of the expansion process occurs at $L_{m_n}$, after
$m_n$ collisions. The subscript $n$ in $w_{0,n}$, $m_n$ and other
  quantities is a reminder that this expansion process is the
  continuation of a compression process that had $n$ collisions.
Let $\beta_n$ be the minimum number of possible collisions during this
expansion process. Then
  $m_n=\beta_n,\cdots,\beta_n+M_{e,n}$, where
%\begin{linenomath}
\begin{align}
  L_{m_n}\leq L_r < L_{m_n+1}.
  \label{exp_ineq}
\end{align}
%\end{linenomath}
Using equation (\ref{Lk}) in relation (\ref{exp_ineq}) we obtain
%\begin{linenomath}
\begin{align}
    g(w_{0,n})< m_n\leq g(w_{0,n})+1,
\end{align}
%\end{linenomath}
where
%\begin{linenomath}
\begin{align}
  g(w_{0,n})=\frac{1}{2}\left(1-\frac{L_1}{L_r}\right)\frac{|w_{0,n}|-|u|}{|u|}.
\end{align}
%\end{linenomath}

Proceeding in a similar way to the compression case, we obtain the transition points ($\eta$) for the negative velocity region ($w_0=-|w_0|$)
%\begin{linenomath}
\begin{align}
  \eta_{m_n}^{-}&=(L_r-L_l)\frac{|w_{0,n}|}{|u|}-L_r(2m_n-1)\nonumber\\
  &=(L_r-L_l)\frac{|v_0|}{|u|}-L_r(2(m_n-n)-1)-2L_ln,
  \label{xminus}
\end{align}
%\end{linenomath}
where we have used equation (\ref{viexp}), as well as the points in the positive velocity region ($w_0=|w_0|$)
%\begin{linenomath}
\begin{align}
  \eta_{m_n}^{+}=-(L_r-L_l)\frac{|v_0|}{|u|}+L_r(2(m_n-n)-1)+2L_ln.
  \label{xplus}
\end{align}
%\end{linenomath}
The different valid values for $\eta_{m_n}^{-}$ and $\eta_{m_n}^{+}$ are found in a similar way to the one presented in the compression case, obtaining this time the points at which the increment from $m_n-1$ to $m_n$ occurs.\\

\begin{figure}[h]
  \centering
  \includegraphics[scale=0.232]{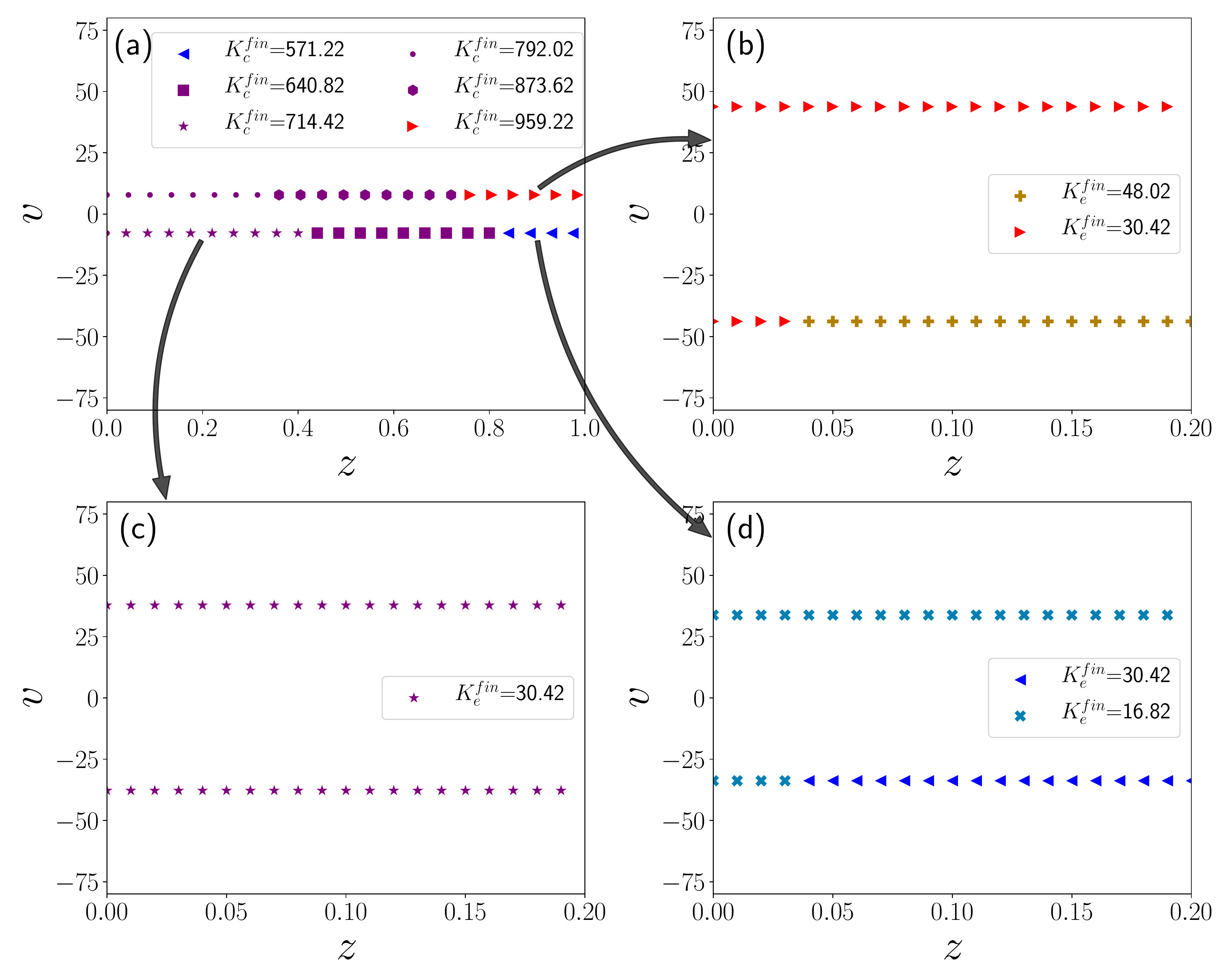}
  \caption{Phase space diagrams for $L_l/L_r=0.2$ and $|v_0|/|u|=7.8$. $K^{ini}$ (in units of $\mu u^2$) for each figure indicate the initial kinetic energy of the particle, while $K^{fin}$ in the legend represent the kinetic energy after the process has ended for the compression (a) or expansion (b to d) processes. (a) $K^{ini}_c=30.42$, (b) $K^{ini}_e=959.22$, (c) $K^{ini}_e=714.42$ and (d) $K^{ini}_e=571.22$. The arrows indicate which expansion process spawns from which region at the start of the compression. The expansion cases not shown in the figure lead all to a final energy equal to the initial one $K^{fin}_e=K^{ini}_c$.}
  \label{phase1}
\end{figure}

\begin{figure}[h]
  \centering
  \includegraphics[scale=0.232]{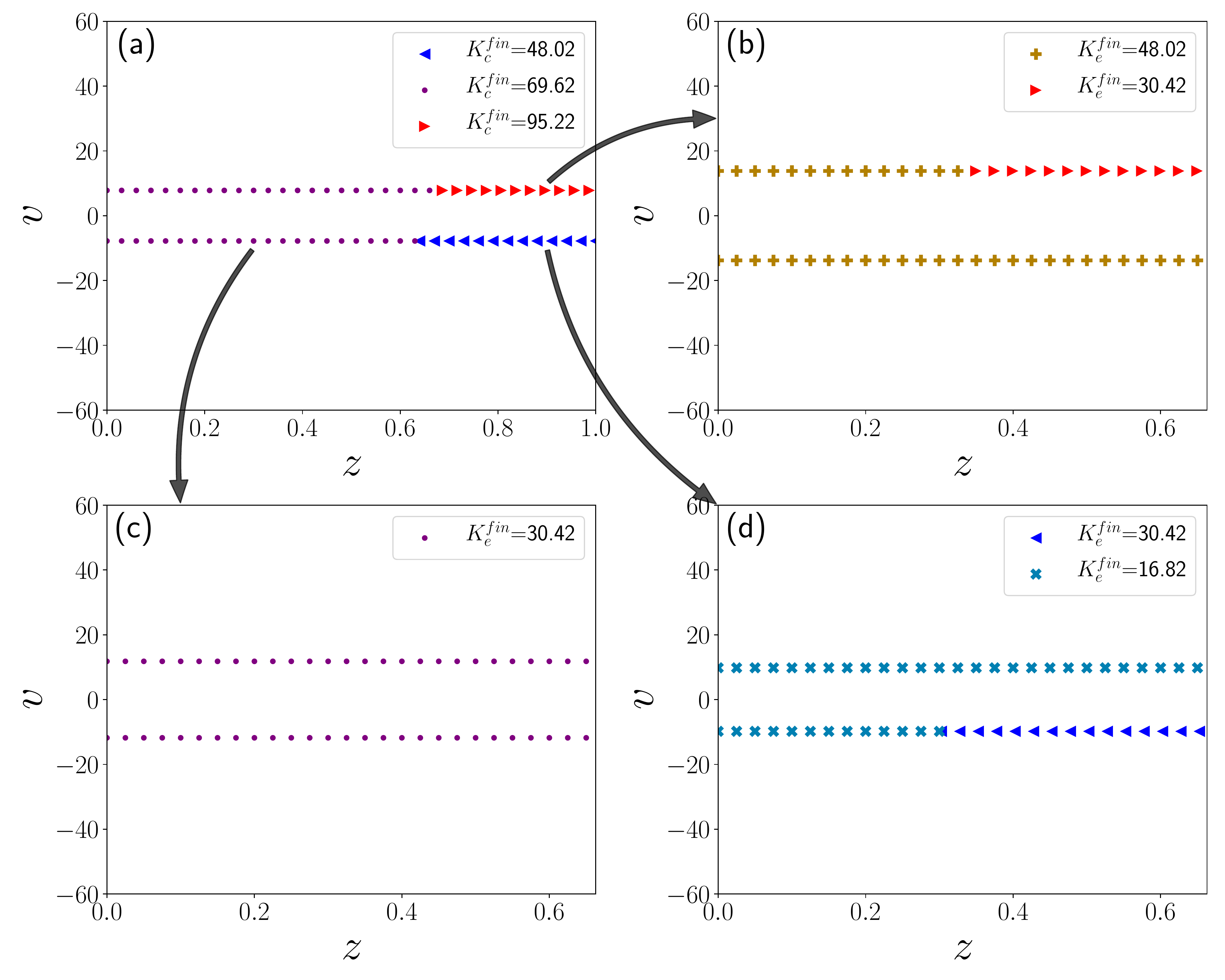}
  \caption{Phase space diagrams for $L_l/L_r=0.663$ and $|v_0|/|u|=7.8$. $K^{ini}$ (in units of $\mu u^2$) for each figure indicate the initial kinetic energy of the particle, while $K^{fin}$ in the legend represent the kinetic energy after the process has ended for the compression (a) or expansion (b to d) processes. (a) $K^{ini}_c=30.42$, (b) $K^{ini}_e=95.22$, (c) $K^{ini}_e=69.62$ and (d) $K^{ini}_e=48.02$. The arrows indicate which expansion process spawns from which region at the start of the compression.}
  \label{phase2}
\end{figure}

Figures \ref{phase1} and \ref{phase2} show the phase diagrams for two processes with $|v_0|/|u|=7.8$ and with left limits for the piston equal to $L_l/L_r=0.2$ and $L_l/L_r=0.663$, respectively.
In the phase space ($z,v$) of the state of the particle at the beginning of the compression we observe several regions, each one corresponding to a different number of collisions during the process.
A symmetry breaking is produced here, as the original phase space volume (length in this case) occupied by the system is separated during the process into $M_c+1$ regions at different energy shells \cite{Parrondo2015}, but unlike systems in contact with a thermal bath \cite{Parrondo2001} it is not thermal fluctuations that choose the actual macroscopic path, as it depends solely on the initial conditions.\\

During the process the size of each region is preserved due to Liouville's theorem. The size ($\mathcal{L}_{c,n}$) of these regions is simply
%\begin{linenomath}
\begin{align}
  \mathcal{L}_{c,n}=|\zeta_{n+1}^{\pm}-\zeta_{n}^{\pm}|,
  \label{lencomp}
\end{align}
%\end{linenomath}
where $\zeta_{\alpha}^{-}=L_r$ and $n=\alpha,\cdots,\alpha+M_c$.\\

Note that sometimes the region at the left side of the container can have one part of it in the negative velocity region and another part in the positive velocity region, as seen in Figure \ref{phase2}(a). For these particular regions
%\begin{linenomath}
\begin{align}
  \mathcal{L}_{c,n}=|\zeta_{n+1}^{+}-0|+|\zeta_{n}^{-}-0|=\zeta_{n+1}^{+}+\zeta_{n}^{-}.
  \label{lencomp}
\end{align}
%\end{linenomath}

At the end of the compression the piston stops for a certain amount of time. During this time the system evolves with a time independent Hamiltonian, which may produce a loss of information \cite{Parrondo2015} in the cases where the size of the constant energy surface, under the time independent Hamiltonian, is larger than that of the region containing the original trajectories of the system, just before the piston stopped. This is evidenced in the plots of the phase space at the beginning of the expansion process by regions that lead to a final energy for the expansion ($K_e^{fin}$) different from the energy at the start of the compression ($K_c^{ini}$).\\

Each of the plots of the phase space at the beginning of expansion in Figs. \ref{phase1}(b to d) and \ref{phase2}(b to d) represents a different energy shell at the end of the trajectory of the compression process, which can also be divided in several regions. In this case the size ($\mathcal{L}_{e,m_n}$) of region $m_n$, given that the compression started at region $n$, is
%\begin{linenomath}
\begin{align}
  \mathcal{L}_{e,m_n}=|\eta_{m_n+1}^{\pm}-\eta_{m_n}^{\pm}|,
  \label{lenexp}
\end{align}
%\end{linenomath}
with $\eta_{\beta_n}^{-}=L_l$  and $m_n=\beta_n,\cdots,\beta_n+M_{e,n}$, where there are $M_{e,n}+1$ regions in the phase space at the beginning of the expansion, given that the compression started at region $n$. In the case in which a region has a part in the positive velocity region and a part in the negative one
%\begin{linenomath}
\begin{align}
  \mathcal{L}_{e,m_n}=|\eta_{m_n+1}^{+}-0|+|\eta_{m_n}^{-}-0|=\eta_{m_n+1}^{+}+\eta_{m_n}^{-}.
  \label{lencomp}
\end{align}
%\end{linenomath}

It is observed that the cases in which $K^{fin}_e$ is different from $K^{ini}_c$, and work different from zero is performed, are associated with starting regions of size $\mathcal{L}_{c,n}<2L_l/L_r$. Table \ref{tableA} shows, for the compression process, the number $\mbox{N}_1$ of regions with size $\mathcal{L}_{c,n}=2L_l/L_r$, the number $\mbox{N}_2=M_c+1-\mbox{N}_1$ of regions with size $\mathcal{L}_{c,n}<2L_l/L_r$, and the combined size
%\begin{linenomath}
\begin{align}
  \mathcal{L}_c^{rem}&=\sum_{n}\mathcal{L}_{c,n},\hspace{0.5cm}\mbox{for }n\mbox{ s.t. } \mathcal{L}_{c,n}<2L_l/L_r\nonumber\\
  &=2(1-\mbox{N}_1L_l/L_r)
\end{align}
%\end{linenomath}
of this last group of regions, for several values of $L_l/L_r$ and $|v_0|/|u|$.\\
\begin{table}[h]
  \centering
  \caption{Number of regions of size $\mathcal{L}_{c,n}=2L_l/L_r$ ($\mbox{N}_1$), $\mathcal{L}_{c,n}<2L_l/L_r$ ($\mbox{N}_2$), and the combined size ($\mathcal{L}_c^{rem}$) for various combinations of the parameters $L_l/L_r$ and $|v_0|/|u|$.}
  \begin{tabularx}{0.45\textwidth}{YYYYY}
    \hline
    \hline
    $L_l/L_r$ & $|v_0|/|u|$ & $\mbox{N}_1$ & $\mbox{N}_2$ & $\mathcal{L}_c^{rem}$\\
    \hlineB{1.5}
    \multirow{3}{*}{0.2} & 7.75 & 4 & 2 & 0.4\\
    & 7.8 & 4 & 2 & 0.4\\
    & 8 & 5 & 0 & 0\\
    \hline
    \multirow{3}{*}{0.45} & 7.5 & 2 & 2 & 0.2\\
    & 7.8 & 1 & 2 & 1.1\\
    & 8.2 & 1 & 2 & 1.1\\
    \hline
    \multirow{3}{*}{0.663} & 7.8 & 1 & 2 & 0.674\\
    & 8.5 & 1 & 2 & 0.674\\
    & 9 & 0 & 2 & 2\\
    \hline
    \hline
  \end{tabularx}
  \label{tableA}
\end{table}

As stated earlier, a region indexed by $n$ will conserve its size during the compression process due to Liouville's theorem. This means that if $\mathcal{L}_{c,n}=2L_l/L_r$, there will be a single region at the end of the compression and thus the expansion will lead back to the original energy surface ($K^{fin}_e=K^{ini}_c$), that is, any random point within the region will lead back to the same region after the compression and expansion. In the cases where $\mathcal{L}_c^{rem}\neq 0$, the two regions $n=\alpha$ and $n=\alpha+M_c$, located next to the piston in, respectively, the negative and positive velocity parts of phase space, will have sizes $\mathcal{L}_{c,n}<2L_l/L_r$.
This means that whenever the compression process starts from one of these two regions, the phase space volume occupied by the particle will expand at the end, where the system is left to evolve under a time independent Hamiltonian, and two regions will appear at the beginning of the expansion: One of them is of the size and location of the original region but with the velocities inverted, and it will lead the system back to the original energy surface. The other one will lead the system to a new surface for which $K^{fin}_e\neq K^{ini}_c$. In this case a net work will be performed on the particle.\\

\section{Average work}\label{Work}
\label{sec:work}

\begin{figure}[h]
  \centering
  \includegraphics[scale=0.235]{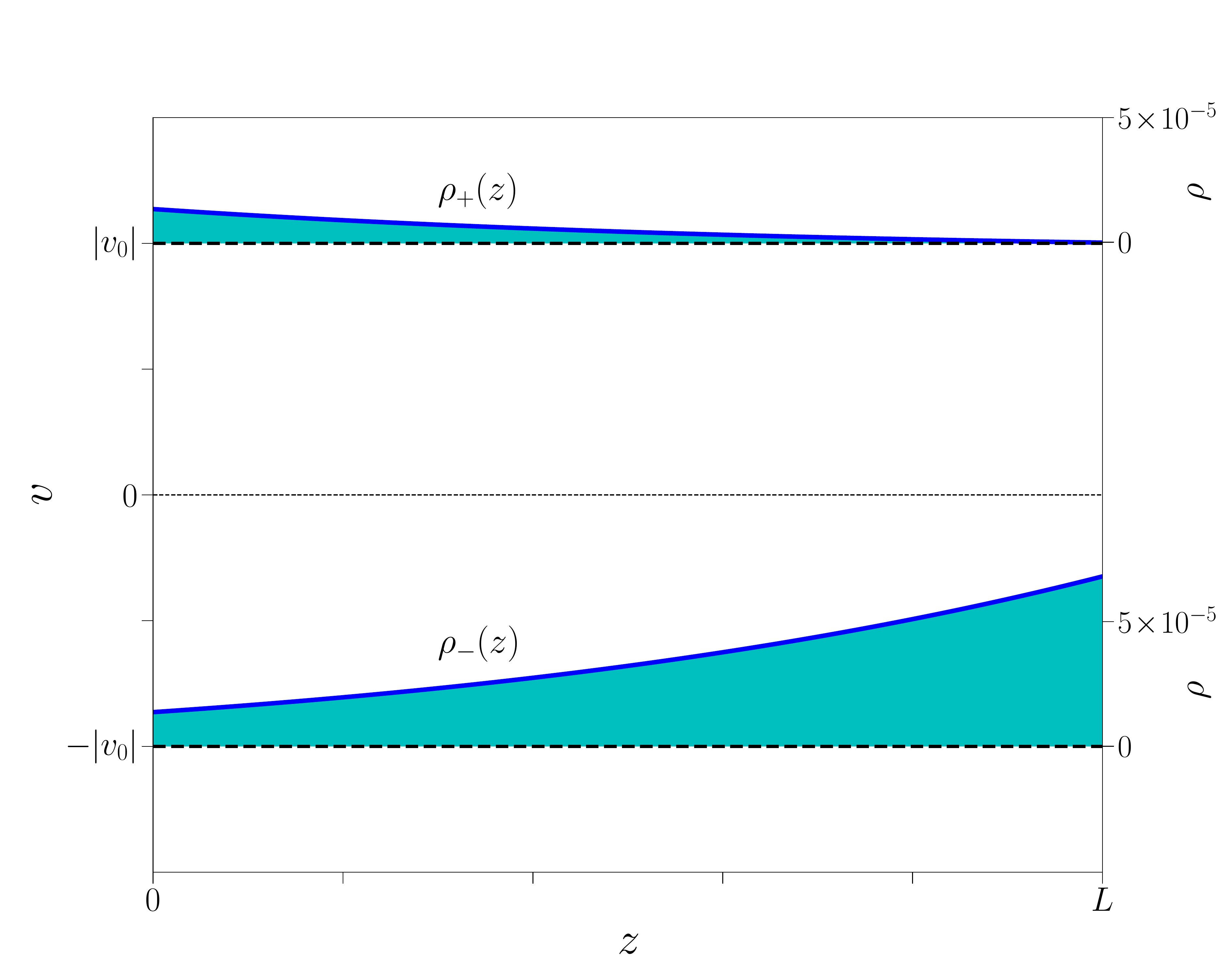}
  \caption{Example of a distribution function of the microstates ($\rho(z,v)$) defined between $0$ and $L$ in each of the regions $v=|v_0|$ and $v=-|v_0|$. $\rho$ has a part on the negative velocity region ($\rho_-(z)$) and a part on the positive velocity region ($\rho_+(z)$). In this example, it is more probable to find a particle initially moving to the left and close to the piston.}
  \label{denspm}
\end{figure}

In this section we will compute the average work done by an ensemble of identical systems, assuming that they begin the process in the same macrostate but in a microstate which is randomly distributed with distribution
%\begin{linenomath}
\begin{align}
  \rho(z,v)=\rho_+(z)\delta(v-|v_0|)+\rho_-(z)\delta(v+|v_0|),
  \label{inidist}
\end{align}
%\end{linenomath}
where $|v_0|=\sqrt{2K_c^{ini}/\mu}$ is the initial speed of the particle, and $\rho_+(z)$ and $\rho_-(z)$ represent the spatial distribution function in the region of positive and negative velocity, respectively (see Fig. \ref{denspm} for a graphical example).\\

The work done during a particular realization of the process is computed by summing the contribution of each of the $n$ collisions between the particle and the piston\cite{Lua2005}
%\begin{linenomath}
\begin{align}
  W=&\sum_{{i}=0}^{n-1}2\mu\ u(u-|v_i|)\nonumber\\
  =&\ 2\mu(u^2n^2-u|v_0|n)
\end{align}
%\end{linenomath}
where equation (\ref{spabs}) was used.
Therefore the net work done on the particle during a compression process with $n$ collisions is
%\begin{linenomath}
\begin{align}
  W_{c,n}=2\mu(u^2n^2+|u||v_0|n)
  \label{microworkc}
\end{align}
%\end{linenomath}
and the work done on the corresponding expansion process with $m_n$ collisions is
%\begin{linenomath}
\begin{align}
  W_{e,m_n}=2\mu(u^2m_n^2-|u||w_{0,n}|m_n),
  \label{microworke}
\end{align}
%\end{linenomath}
where the particle starts the compression with speed $|v_0|$, and the expansion with speed $|w_{0,n}|$ given by (\ref{viexp}).\\ 

We can use the phase diagrams presented in the previous section to compute the average work done by the ensemble during a compression/expansion process, as they show the dependence of the
work done on the particle on its initial state.
As there is no exchange of heat with an external system during the process, the work done is uniquely determined by the initial state of the particle for a given protocol.
In the case of the compression the work is given by
\begin{widetext}
%\begin{linenomath}
\begin{align}
  W_{c}^{tot}(z,v)=\left\{ \begin{array}{ll}
    \sum_{n=\alpha}^{\alpha+l-1}W_{c,n}\mathbbm{1}(z,\zeta_{n}^{-},\zeta_{n-1}^{-})+W_{c,\alpha+l}\mathbbm{1}(z,0,\zeta_{\alpha+l}^{-}), & \mbox{if }v<0\\
    \sum_{n=\alpha+l+1}^{\alpha+M_c}W_{c,n}\mathbbm{1}(z,\zeta_{n-1}^{+},\zeta_{n}^{+})+W_{c,\alpha+l}\mathbbm{1}(z,0,\zeta_{\alpha+l+1}^{+}), & \mbox{if }v>0\\
  \end{array} \right .,
  \label{compwork}
\end{align}
%\end{linenomath}
\end{widetext}
where $W_{c,n}$, computed as in (\ref{microworkc}),  is the work done during the compression when the particle starts at region $n$,
and
%\begin{linenomath}
\begin{align}
  \mathbbm{1}(z,a,b)=\Theta(z-a)[1-\Theta(z-b)]
\end{align}
%\end{linenomath}
is the characteristic function of the interval $[a,b]$,
with $\Theta(x)$ representing the Heaviside function. 
The index $n$ runs over all the $M_c+1$ regions in phase space, located on the negative velocity region from $n=\alpha$ to $n=\alpha+l$, and on the positive region from $n=\alpha+l+1$ to $n=\alpha+M_c$.
In general, the region at the $z=0$ border (identified by the index $n=\alpha+l$) may be split in two parts.\\

Similarly, we have that the function giving the work done during the
expansion starting at region $m_n$ (after a compression that started
at region $n$), has the form
\begin{widetext}
%\begin{linenomath}
\begin{align}
  W_{e,n}^{tot}(z,w)=\left\{ \begin{array}{ll}
    \sum_{m_n=\beta_n}^{\beta_n+s-1}W_{e,m_n}\mathbbm{1}(z,\eta_{m_n}^{-},\eta_{m_n-1}^{-})+W_{e,\beta_n+s}\mathbbm{1}(z,0,\eta_{\beta_n+s}^{-}), & \mbox{if }w<0\\
    \sum_{m_n=\beta_n+s+1}^{\beta_n+M_{e,m_n}}W_{e,m_n}\mathbbm{1}(z,\eta_{m_n-1}^{+},\eta_{m_n}^{+})+W_{e,\beta_n+s}\mathbbm{1}(z,0,\eta_{\beta_n+s+1}^{+}), & \mbox{if }w>0\\
  \end{array} \right .,
  \label{expwork}
\end{align}
%\end{linenomath}
\end{widetext}
where $W_{e,m_n}$ is the work done during the expansion when the particle starts at region $m_n$, computed as in (\ref{microworke}).
%\refA{
Regions from $m_n=\beta_n$ to $m_n=\beta_n+s$ are located on the negative velocity part of the phase space, while regions from $m_n=\beta_n+s+1$ to $m_n=\beta_n+M_{e,m_n}$ are located in the positive part. 
The region for which $m_n=\beta_n+s$ is generally split between the positive and negative velocity parts.\\
%}

We can compute now the average work done during the compression using equations (\ref{inidist}) and (\ref{compwork}) as
%\begin{linenomath}
\begin{align}
  \left<W_c^{tot}\right>=&\int_0^{L_r}\int_{-\infty}^{\infty}\ W_c^{tot}(z,v)\rho(z,v)dvdz\nonumber\\
  =&\sum_{n=\alpha}^{\alpha+l-1}a_{n}W_{c,n}+(a_{\alpha+l}+a_{\alpha+l}')W_{c,\alpha+l}\nonumber\\
  &+\sum_{n=\alpha+l+1}^{\alpha+M_c}a_{n}'W_{c,n},
\end{align}
%\end{linenomath}
where
%\begin{linenomath}
\begin{align}
  a_{n<\alpha+l}=&\int_{\zeta_{n+1}^{-}}^{\zeta_{n}^{-}}\rho_-(z)\ dz,\label{coef1}
\end{align}
\begin{align}
  a_{n>\alpha+l}'=&\int_{\zeta_{n}^{+}}^{\zeta_{n+1}^{+}}\rho_+(z)\ dz,
\end{align}
\begin{align}
  a_{\alpha+l}=&\int_{0}^{\zeta_{\alpha+l}^{-}}\rho_-(z)\ dz,\mbox{ and}
\end{align}
\begin{align}
  a_{\alpha+l}'=&\int_{0}^{\zeta_{\alpha+l+1}^{+}}\rho_+(z)\ dz,\label{coef4}
\end{align}
%\end{linenomath}
where region $n=\alpha+l$ receives a special treatment as it lies in both, positive and negative velocity parts of the constant energy shell. If $\rho(z,v)$ is uniform (microcanonical distribution) we will have\\
%\begin{linenomath}
\begin{align}
  \left<W_c^{tot}\right>=&\sum_{n=\alpha}^{\alpha+M_c}a_{n}W_{c,n},
\end{align}
%\end{linenomath}
with
%\begin{linenomath}
\begin{align}
  a_{n}=\frac{\mathcal{L}_{c,n}}{2L_r},
\end{align}
%\end{linenomath}
where $\mathcal{L}_{c,n}$ is defined in (\ref{lencomp}).
The coefficients $a_{n}$ represent the probability of starting the compression process within region $n$, so
%\begin{linenomath}
\begin{align}
  \sum_{n=\alpha}^{\alpha+M_c}a_{n}=1.
\end{align}
%\end{linenomath}
Assuming now that the probability distribution of the microstates at the beginning of the expansion is always uniform, we obtain that the average work done during the expansion is given by
%\begin{linenomath}
\begin{align}
  \left<W_e^{tot}\right>=&\sum_{n=\alpha}^{\alpha+M_c}\frac{1}{2L_l}\int_0^{L_l}\int_{-\infty}^{\infty}\ W_{e,n}^{tot}(z,w)\Big[\delta(w-|w_{0,n}|)\nonumber\\
    &+\delta(w+|w_{0,n}|)\Big]dwdz\nonumber\\
  =&\sum_{n=\alpha}^{\alpha+M_c}\left(\sum_{m_n=\beta_n}^{\beta_n+M_{e,n}}b_{m_n}W_{e,m_n}\right),
\end{align}
%\end{linenomath}
with
%\begin{linenomath}
\begin{align}
  b_{m_n}=\frac{\mathcal{L}_{e,m_n}}{2L_l},\label{coefc}
\end{align}
%\end{linenomath}
where $\mathcal{L}_{e,m_n}$ is defined in (\ref{lenexp}).
The coefficients $b_{m_n}$ represent the conditional probability of starting the expansion within region $m_n$ given that the compression started at $n$, and thus
%\begin{linenomath}
\begin{align}
  \sum_{m_n=\beta_n}^{\beta_n+M_{e,n}}b_{m_n}=1.
\end{align}
%\end{linenomath}

Finally, the average work done during both the compression and expansion for a particular value of the parameters $L_r$, $L_l$, $|u|$ and $|v_0|$ is
%\begin{linenomath}
\begin{align}
  \left<W\right>=\left<W_c^{tot}\right>+\left<W_e^{tot}\right>.
\end{align}\\
%\end{linenomath}

As seen before, the trajectories that start the compression at regions in phase space which are adjacent to the piston have the potential, after the reverse process (expansion) has been carried out,
to lead the system into
a macroscopic final state which is different from the initial one, thus performing a net work on the particle.\\

We compute the probability of doing net positive work on the particle ($p_{pos}$) as
%\begin{linenomath}
\begin{equation}
  p_{pos}=a_{\nu}b_{\beta_{\nu}},
\end{equation}
%\end{linenomath}
where $a_{\nu}$ is the probability of starting the compression in the region located next to the piston in the positive velocity part of the phase space ($\nu=\alpha+M_c$), and $b_{\beta_{\nu}}$ is the probability of starting the expansion in the region adjacent to the piston with negative velocity, given that it started the compression at region $\nu$. Likewise, the probability of doing net negative work ($p_{neg}$) is
%\begin{linenomath}
\begin{equation}
  p_{neg}=a_{\alpha}b_{\omega_{\alpha}},
\end{equation}
%\end{linenomath}
where $a_{\alpha}$ is the probability of starting the compression in the region located next to the piston in the negative velocity part of the phase space, and $b_{\omega_{\alpha}}$ is the probability of starting the expansion in the region adjacent to the piston with positive velocity, given that it started the compression at region $\alpha$ ($\omega_{\alpha}=\beta_{\alpha}+M_{e,\alpha}$).
Finally, the probability for the process to do no net work is
%\begin{linenomath}
\begin{align}
  p_0=1-p_{pos}-p_{neg}.
\end{align}
%\end{linenomath}

\begin{figure}[h]
  \centering
  \includegraphics[scale=0.085]{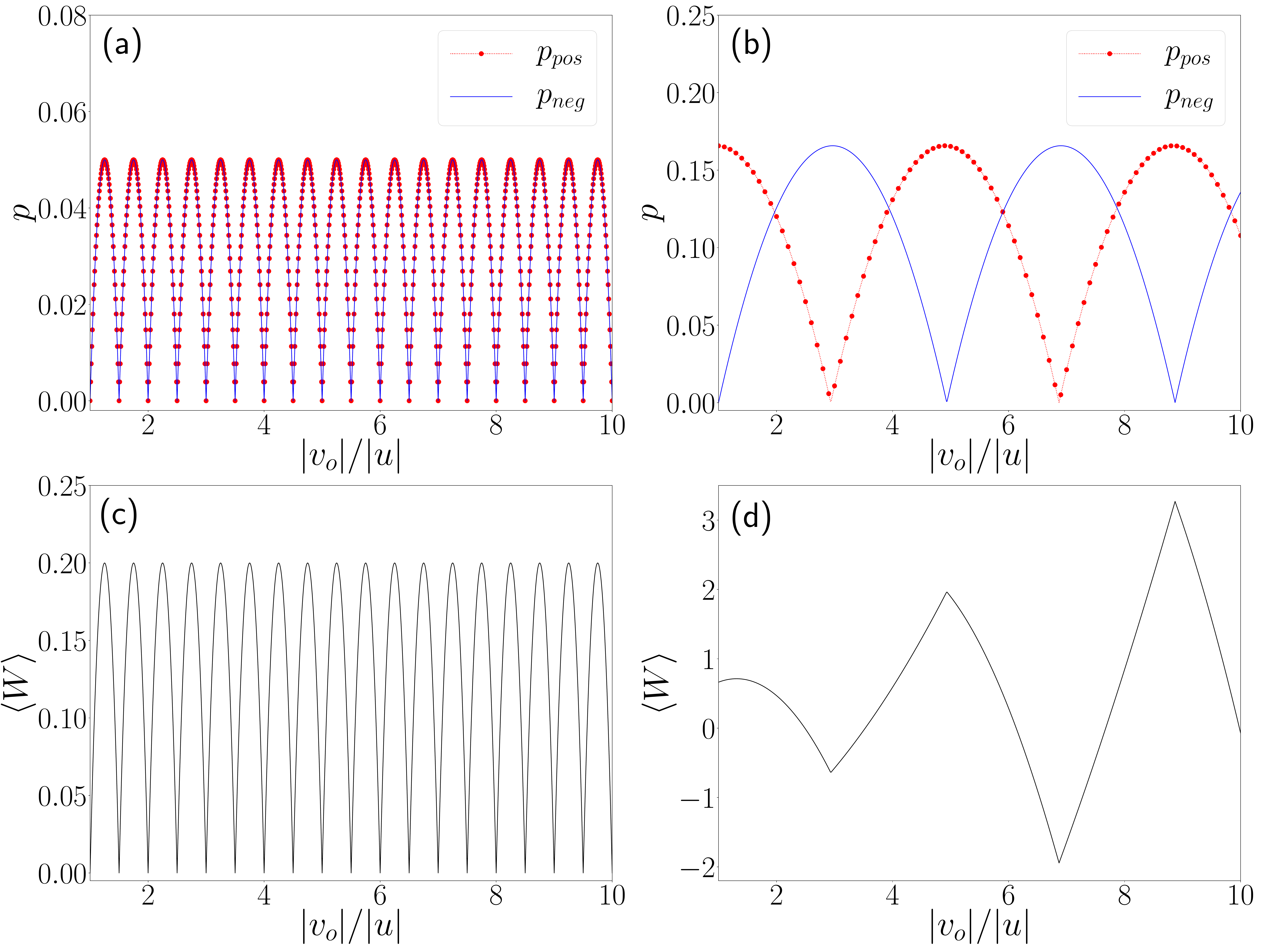}
  \caption{Upper figures: Probability of doing positive $p_{pos}$ (dotted line) or negative $p_{neg}$ (continuous line) work on the particle for a system with (a) $L_l/L_r=0.2$ or (b) $L_l/L_r=0.663$. Lower figures: Average work done on the particle for (c) $L_l/L_r=0.2$ or (d) $L_l/L_r=0.663$.}
  \label{avwork}
\end{figure}

Figures \ref{avwork}(a,b) show the probabilities of doing positive and negative work as a function of the initial speed ($|v_0|/|u|$) of the particle, assuming that the initial distribution of microstates is uniform. It is observed that the probability of doing net work on the particle different from zero has a periodic behavior. Furthermore, depending on the parameter $L_l$ the graphs for $p_{pos}$ and $p_{neg}$ can be in phase (Fig. \ref{avwork}(a)), where the probability for the process to do net work falls to zero at some points, or out of phase (Fig. \ref{avwork}(b)), where there is always a possibility for the process to do work different from zero. In this last case there are regions in which it is more probable to do positive rather than negative work on the particle and vice versa.\\    

The average work as a function of the speed is shown in Figs. \ref{avwork}(c) and (d). In the case in which the probabilities $p_{pos}$ and $p_{neg}$ are always equal, no negative work is observed at any initial speed (Fig. \ref{avwork}(c)). To understand this, we note that in this system the thermodynamic irreversibility is produced by the difference $d=m_n-n$ between the number of hits during the compression and its inverse process, the expansion. If $d<0$ the net work is positive, while $d>0$ leads to a net negative work. However, the magnitude of the energy transfer produced by a collision during the compression, starting with the particle moving to the right at speed $|v|$ and the piston moving to the left at speed $|u|$, is larger than that of a collision during the expansion, in which both the particle and the piston start moving to the right at those same speeds $|v|$ and $|u|$ respectively. On the other hand, when $p_{pos}$ and $p_{neg}$ are different there is the possibility to do either positive or negative work on average, depending of the initial speed of the particle (Fig. \ref{avwork}(d)). \\

\begin{figure}[h]
  \centering
  \includegraphics[scale=0.085]{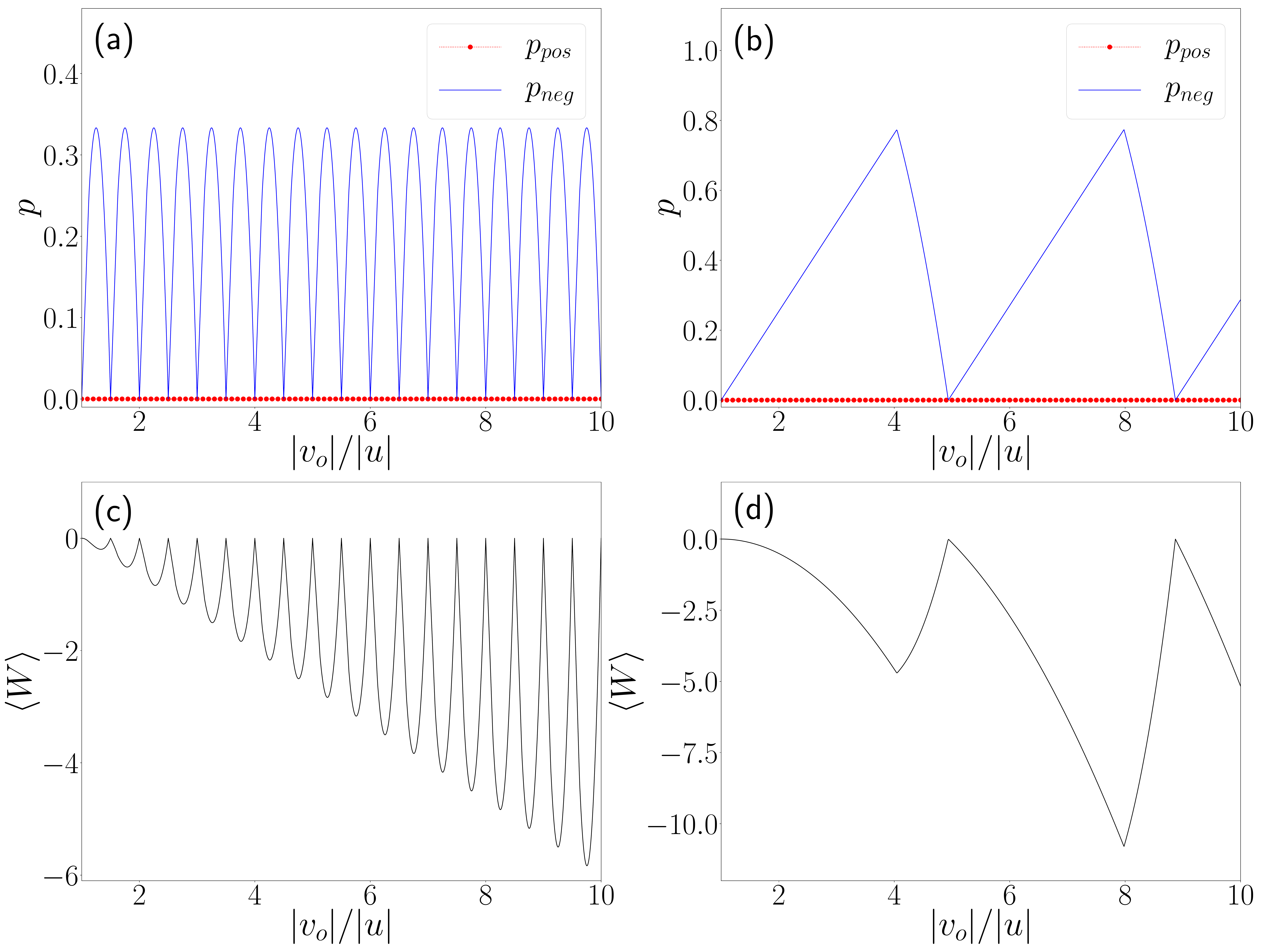}
  \caption{Upper figures: Probability of doing positive $p_{pos}$ (dotted line) or negative $p_{neg}$ (continuous line) work on the particle when using distributions (\ref{distp}) and (\ref{distm}) with $z_{min}/L_r=0.7$, for a system with (a) $L_l/L_r=0.2$ or (b) $L_l/L_r=0.663$. Lower figures: Average work done on the particle for (c) $L_l/L_r=0.2$ or (d) $L_l/L_r=0.663$.}
  \label{avwork_fix}
\end{figure}

We look now at a case in which the initial distribution of states for the particle is not uniform.
Figure \ref{avwork_fix} shows the probabilities $p_{pos}$ and $p_{neg}$, as well as the average work done on the particle when the position dependent parts of the distribution function (\ref{inidist}) are given by
%\begin{linenomath}
\begin{align}
  \rho_+(z)=&\ 0,\label{distp}\\
  \rho_-(z)=&\left\{\begin{array}{ll}
  \frac{1}{(L_r-z_{min})}\Theta(z-z_{min}) & \mbox{, if }0\leq z\leq L_r\\
  0 & \mbox{, otherwise}\label{distm}
  \end{array}\right.
  .
\end{align}
%\end{linenomath}
That is, the particle starts at a region close to the piston at a
distance at most $L_r-z_{min}$ from it, where $z_{min}>L_l$ is a given
parameter (chosen as $z_{min}=0.7 L_r$ in Fig. \ref{avwork_fix}).
It is observed that when the possibility for the particle to start in
the positive velocity region is precluded, the probability of doing
positive average work on the system also vanishes. This means that the
particle either maintains its kinetic energy or work has been extracted by the end of the compression/expansion process.

\section{Irreversibility}
\label{sec:reversibility}

In this section we define a measure of thermodynamic irreversibility for the system.
As we have seen, the phase space diagrams of the initial states in a
compression or expansion process show that the phase space is divided
into regions with different number of collisions. This breaks the
ergodicity of the system, as two different realizations starting the
process at opposite sides of the boundary between two regions would
end up at different energy surfaces. \\ 

As the evolution of the macroscopic state $(K,L)$ of the system
depends on its initial microscopic state $(z,v)$ only to the extent
that it is located within a particular region of the phase space, in
the following we will drop the microscopic state of the particle as a
variable and use instead the indexes $n$ and $m_n$ to denote the
region in which it is initially located. As the evolution of the
system during a compression or an expansion is Hamiltonian, the
distribution of states follows from Liouville's equation and the
trajectory followed by the system depends exclusively on its initial
state.
However, between the compression and the expansion, a
  relaxation process is introduced where the piston remains
  motionless. During this time, the system is allowed to relax to a
  combination of microcanonical distributions, i.e. for each of the
  final energies obtained after the compression, the corresponding
  energy shell is allowed to be uniformly distributed in phase space
  before starting the expansion process. This relaxation process
  introduces irreversibility in the system which, in the following, we
  propose a way to quantify it. We start with the analogue situation for systems
  in contact with a thermal reservoir.\\

  For a system which is allowed to equilibrate with a thermal bath at temperature $T$ at the beginning and at the end of a process, the Crooks relation
%\begin{linenomath}
\begin{align}
  \frac{\mathcal{P}^F[\gamma_F]}{\mathcal{P}^R[\gamma_R]}=e^{(W_F-\Delta F)/k_BT}
\end{align}
%\end{linenomath}
is obtained\cite{Crooks1998, Jarzynski2011}, where
$\mathcal{P}^F[\gamma_F]$ is the probability distribution for
obtaining a particular forward path $\gamma_F$ (during which a
  work $W_F$ is performed on the system) and
$\mathcal{P}^R[\gamma_R]$ is the probability distribution for
obtaining its reverse path $\gamma_R$. $\Delta F$ is the
  equilibrium free
  energy difference between the final and initial states of the
  process. The logarithm of this relation has been used as a measure of the irreversibility of the path $\gamma$ leading from an initial point in phase space $\bm{x}_{i}$ to a final point $\bm{x}_{f}$\cite{Spinney2013}
%\begin{linenomath}
\begin{align}
  I[\gamma_F]=\ln\left[\frac{\mathcal{P}^F[\gamma_F]}{\mathcal{P}^R[\gamma_R]}\right].
  \label{irrT}
\end{align}
%\end{linenomath}
We note here that the process under these conditions does not present a breaking of ergodicity, as the thermalization at both ends of the process makes the whole phase space accessible to the system.\\

In our case, we define a similar measure of irreversibility for the path followed by the system when it starts the compression at region $n$ as
%\begin{linenomath}
\begin{align}
  I_{n}=&\ln\left[\frac{P_{n}^F}{P_{\bar{n}}^R}\right],
  \label{irr}
\end{align}
%\end{linenomath}
where $P_{n}^F$ is the probability for the system to start the compression at region $n$ and $P_{\bar{n}}^R$ is the probability for the system to start the expansion at region $\bar{n}$, which is the region in phase space at which the system arrived at the end of the compression, but with all momenta reversed. In this case, because of Liouville's theorem, the system goes through the reverse path leading back to region $n$. \\ 

The probability $P_{n}^F$ is given by
%\begin{linenomath}
\begin{align}
  P_{n}^F=a_{n},
  \label{forwardp}
\end{align}
%\end{linenomath}
where the coefficients $a_{n}$ are given by (\ref{coef1}) to (\ref{coef4}).
On a similar manner, the conditional probability for the particle to be within region $m_n$ at the start of the expansion, given that it started the compression at region $n$ is given by
%\begin{linenomath}
\begin{align}
  P_{m_n}^R=b_{m_n},
\end{align}
%\end{linenomath}
where the coefficients $b_{m_n}$ are given by (\ref{coefc}).\\

Let $m$ denote a region in phase space at the start of the expansion process, irrespective of the region at which the compression started, and $P_{m}^R$ the probability for the system to start the expansion at this region. Then
%\begin{linenomath}
\begin{align}
  P_{m}^R=\sum_{k=\alpha}^{\alpha+M_c}P_{m_k}^RP_k^F=\sum_{k=\alpha}^{\alpha+M_c}b_{m_k}a_{k}.
\end{align}
%\end{linenomath}
However, as the ergodicity is broken during the compression we know that region $m$ is accessible only to processes that started at a certain region that we call $n$. Therefore $b_{m_k}\neq 0$ only if $k=n$, and
%\begin{linenomath}
\begin{align}
  P_{m}^R=b_{m_n}a_{n},
\end{align}
%\end{linenomath}
for $m_n=\beta_n,\cdots,\beta_n+M_{e,n}$. This means that the probability of following the reverse path is
%\begin{linenomath}
\begin{align}
  P_{\bar{n}}^R=b_{\bar{n}_n}a_{n}.
  \label{reversp}
\end{align}
%\end{linenomath}

\begin{figure}
  \includegraphics[scale=0.085]{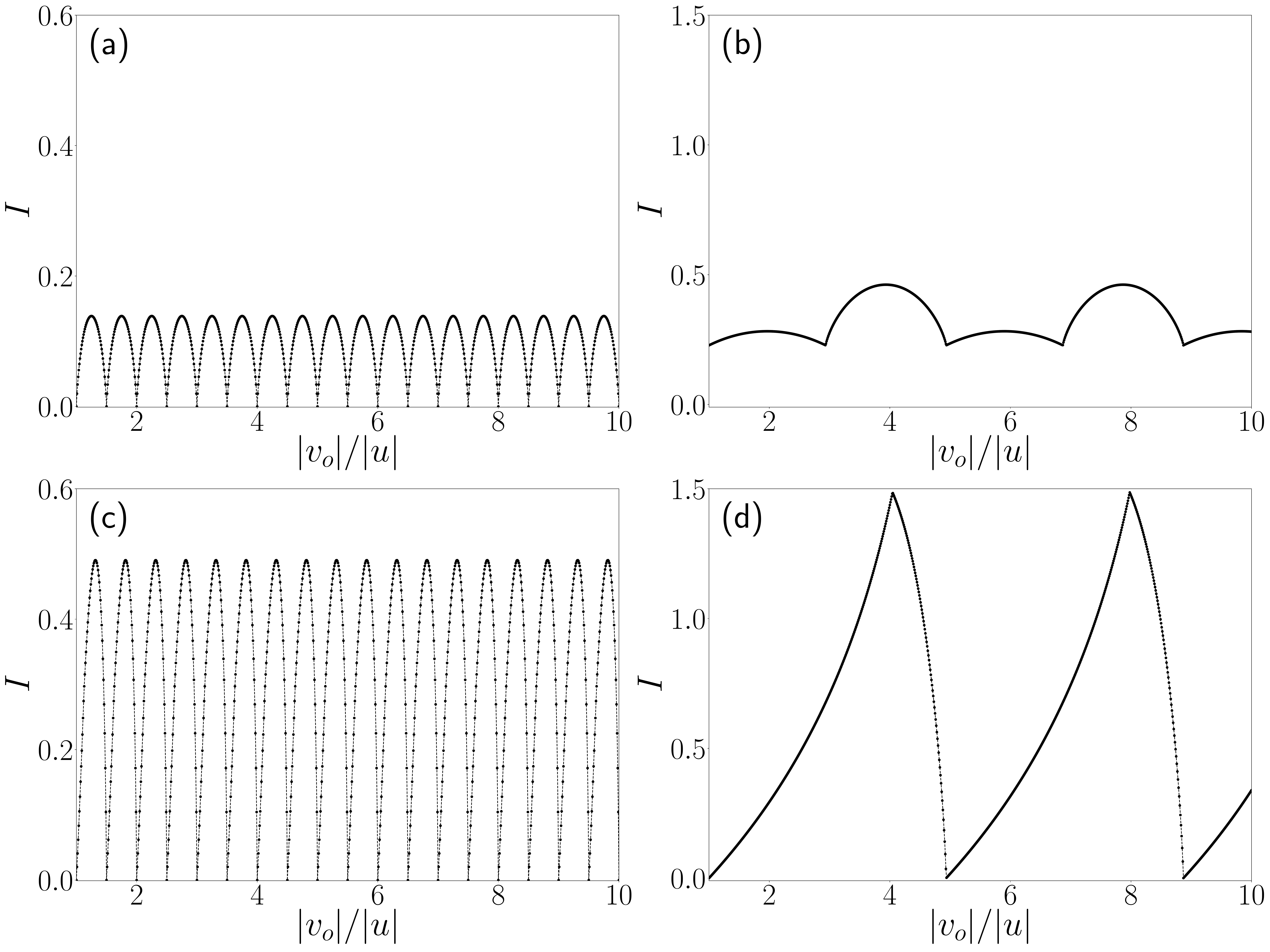}
  \vspace{-0.5cm}
  \caption{Irreversibility, as defined by equation (\ref{avirr}) for systems starting with microcanonical distribution (a) $L_l/L_r=0.2$ and (b) $L_l/L_r=0.663$, as well as distribution (\ref{distp})-(\ref{distm}) with $z_{min}/L_r=0.7$ for (c) $L_l/L_r=0.2$ and (d) $L_l/L_r=0.663$.}
  \label{irrevfix}
\end{figure}

Using equations (\ref{forwardp}) and (\ref{reversp}) into equation (\ref{irr}) we obtain
%\begin{linenomath}
\begin{align}
  I_{n}=\ln\left[\frac{1}{b_{\bar{n}_n}}\right]\geq0.
\end{align}
%\end{linenomath}
Finally, the average irreversibility over all the starting regions $n$ is
%\begin{linenomath}
\begin{align}
  I=\sum_{n=\alpha}^{\alpha+M_c}a_{n}I_{n}.
  \label{avirr}
\end{align}
%\end{linenomath}
Figure \ref{irrevfix} plots $I$ as a function of $|v_0|/|u|$ for two values of $L_l$ in the case in which the system starts from a microcanonical distribution of states and also in the case in which the distribution is the step function given by (\ref{distp}) and (\ref{distm}). 
It is observed that the system is reversible ($I=0$) whenever the net work done on the particle is zero (see Figs. \ref{avwork} and \ref{avwork_fix}), and that it reaches some maximum values periodically, representing points at which the probability for the system to take the reverse path during the expansion, back to the starting region in phase space is small compared to the probability of taking the corresponding forward path during the compression. The value of such maxima increases considerably when we restrict the initial states to the region of negative velocities close to the piston.

\section{Summary and Conclusions}
\label{sec:conclusions}

  In this work we have studied the conditions under which the work done on a single particle by a piston moving at constant speed, and which performs a compression followed by an expansion, is different from zero. This was done by obtaining an analytical result relating the number of hits $n$ between the piston and the particle, with the position at which the $n$th collision occurs. Using this relation, diagrams in the phase space of initial conditions $(z,v_0)$ were constructed for the particle, by looking for the boundaries between regions with $n$ and $n+1$ collisions during the compression or expansion processes.\\  

  An important aspect of the protocol lies in the relaxation time that takes place between the end of the compression and the beginning of the expansion. If an ensemble of identical systems with random starting particle positions is taken, all systems starting the compression at a region of size $\mathcal{L}$ where $n$ collisions occur will, by Liouville's theorem, evolve along with the region towards the same final energy surface, that is different for every region (ergodicity breaking). Once the piston stops, the region in phase space occupied by this subset of systems changes its size to a new value $\mathcal{L}'\geq \mathcal{L}$, as the distribution relaxes to the microcanonical distribution (fixed speed and a uniformly distributed position). In the cases where the size of the region changes, a system might end the expansion with an energy different from the initial one, and in this case the work done on the system is different from zero.\\

  The relaxation time is part of the protocol, and it is important to
  remember that during this lapse the speed of the piston is
  zero. This relaxation is at the heart of the irreversibility of
    the process leading to the possibility of obtaining a non zero
    work. If we want to restore {\it mechanical reversibility}
    not only the momenta of the particle and the piston have to be
  reversed, but also the protocol. This means that if the reversal in
  momenta is carried out during the relaxation phase, when a time
  $\Delta t'$ has passed since the end of the compression, the same
  amount of time $\Delta t'$ has to pass before the beginning of the
  expansion. \\
  
  The phase diagrams show that in order to do net work different from zero, the particle has to start the compression and expansion located in a region adjacent to the piston. Additionally, for this work to be positive, the particle first has to start the compression on the region of positive velocities (moving towards the piston), and once the compression ends and the particle is allowed to equilibrate to the microcanonical distribution, the particle has to start the expansion in the region of negative velocities (moving away from the piston). In a similar manner, for the net work to be negative the particle has to to start the compression on the region of negative velocities, and the expansion in the region of positive velocities. In these cases the number of collisions during the compression ($n$) and the number of collisions during the expansion ($m$) are different, which produces a net change in the energy of the particle, making the process {\it thermodynamically irreversible}.\\
  
  If the process starts in the microcanonical ensemble, the probabilities of doing net positive ($p_{pos}$) or negative ($p_{neg}$) work on the particle as a function of the particle's speed ($|v|/|u|$) during a compression/expansion cycle oscillate between zero and some maximum value. Depending on the parameter $L_l$, these oscillations might be in phase ($p_{pos}=p_{neg}$ always) or not. If they are in phase the work cannot be negative. This happens as, for given values of the speeds of the particle ($|v|$) and the piston ($|u|$), the work done on a collision during the compression transfers a greater amount of energy than a collision during the expansion. On the other hand, if the oscillations of the probabilities are out of phase, there will be values of $|v|/|u|$ for which it is more probable to do either negative or positive work.\\
  
  If the starting distribution of the systems is different from the
  microcanonical one, the amplitude of the oscillations of the
  probabilities $p_{pos}$ or negative $p_{neg}$ might be changed. In
  particular, if the spatial distribution is a step function like the
  one in equations (\ref{distp}) and (\ref{distm}), it is possible to
  preclude the possibility of doing positive work. This implies that
  by performing successive cycles energy can be extracted from the
  particle while its initial speed $|v|$ is greater than
  $|u|$. However, this amounts to a forced symmetry
  breaking\cite{Parrondo2001} and in order to do a complete assessment
  of the net work extracted one should also compute the amount of
    work required to prepare the initial nonuniform distribution from
    the microcanonical ensemble.\\
  
  Finally, regarding quasistatic processes ($|u|$ very small), one can see that even if the work done on a single collision (equation (\ref{work1})), and thus the net work done on the particle during the process, becomes very small, the value of $|d|=|m-n|$ remains an integer which might be different from zero, and the process cannot be said to be {\it Carnot reversible} in general, with the exception of some particular cases for which the irreversibility $I=0$ (see for example Fig. \ref{irrevfix}).\\ 

\section*{Acknowledgments}
GT acknowledges partial support from Fondo de Investigaciones 2018-2019, Facultad de Ciencias, Uniandes.

\bibliographystyle{prsty}

\end{document}